\documentclass[useAMS,usenatbib]{mn2e}
\usepackage{graphicx}

\newcommand{\apj}{ApJ}

\newcommand{\aap}{A\&A}
\newcommand{\aaps}{A\&AS}
\newcommand{\aj}{AJ}
\newcommand{\mnras}{MNRAS}

\newcommand{\pasp}{PASP}
\newcommand{\apjs}{ApJS}
\newcommand{\MC}{\multicolumn}
\newcommand{\kms}{km\,s$^{-1}$}
\newcommand{\ion}[2]{#1\,\textsc{#2}}
\newcommand{\IC}{IC\,10}

\title[Spectroscopy of two PN candidates in \IC]{Spectroscopy of two PN candidates in \IC\thanks{Based on
observations obtained at the 6m~SAO RAS telescope.}
}

\author[A.\ Y. Kniazev, S.\ A. Pustilnik, \& D.\ B. Zucker]{%
A.Y. Kniazev,$^{1,2}$\thanks{E-mail: akniazev@saao.ac.za (AYK); sap@sao.ru (SAP); zucker@ast.cam.ac.uk (DBZ)}
S.A. Pustilnik,$^{2}$
D.B. Zucker$^{3,4}$\\
$^{1}$South African Astronomical Observatory, Observatory Road, Cape Town, South Africa\\
$^{2}$Special Astrophysical Observatory, Nizhnij Arkhyz, Karachai-Circassia,
369167, Russia \\
$^{3}$Max-Planck-Institut f\"{u}r Astronomie, K\"{o}nigstuhl 17,
      D-69117 Heidelberg, Germany \\
$^{4}$Institute of Astronomy, University of Cambridge, Madingley Road, Cambridge CB3 0HA, United Kingdom
}

\begin{document}

\date{Accepted 2007 March ??. Received 2007 January ??; in original form 20?? October ??}

\pagerange{\pageref{firstpage}--\pageref{lastpage}} \pubyear{2007}

\maketitle

\label{firstpage}

\begin{abstract}
We present the results of the first spectroscopic observations of two
planetary nebula (PN) candidates in the Local Group dwarf irregular galaxy
\IC. Using several spectral classification diagrams
we show that the brightest PN candidate (PN\,7)
is not a PN, but rather a compact \ion{H}{ii} region consisting of two components
with low electron number densities.
After the rejection of this PN candidate,
the \IC\ planetary nebula luminosity function cutoff becomes very close
to the standard value.
With the compiled spectroscopic data for a large number
of extragalactic PNe,
we analyse a series of diagnostic diagrams to generate
quantitative criteria for separating PNe from unresolved
\ion{H}{ii} regions.
We show that, with the help
of the diagnostic diagrams and the derived set of
criteria, PNe can be distinguished from
\ion{H}{ii} regions
with an efficiency of $\sim$99.6\%. With
the obtained spectroscopic data we confirm that
another, 1\fm7 fainter PN candidate (PN\,9) is a genuine PN.
We argue that, based on all currently available PNe data, \IC\ is located
at a distance 725$^{+63}_{-33}$ kpc
(distance modulus $\rm (m-M) = 24.30^{+0.18}_{-0.10}$).
\end{abstract}


\section{Introduction}

The Local Group is an excellent laboratory for studies of galaxy evolution, providing
 us with a wide range of different galaxy types in
a variety of environments. The star formation (SF) histories of
Local Group galaxies can be obtained from color-magnitude diagrams
of resolved stars \citep[e.g.,][]{AG05}.
Using this method it is possible to constrain the entire SF and chemical
enrichment history of galaxies.
However, the results are model-dependent, and should be compared
with other, additional observational data which can be obtained
for Local Group galaxies. One such complementary means
of determining the SF and chemical evolution histories is a study of planetary nebulae (PNe),
which can be used simultaneously as age, kinematics and metallicity
tracers.  The abundances from both \ion{H}{ii} regions and PNe allow one to derive
an approximate enrichment history for a galaxy from intermediate ages
to the  present day and permit abundance measurements at different locations.
The latter, in turn, helps to test the homogeneity with which
elements are distributed through a galaxy and, thus, the timescale
for heavy element  diffusion and mixing \citep{Sextans, KGPP04, SextB}.

In this paper, we present new results of a PN study in the Local Group
dwarf irregular galaxy \IC, which is situated at low Galactic
latitude ($b = -3\hbox{$.\!\!^\circ$ }3$).
Estimates of the distance to \IC\ vary, including values of 830$\pm$110 kpc,
based on observations of Cepheids \citep{Sah96};
660$\pm$66 kpc, based on Cepheids and tip of the
red giant branch (RGB) stars \citep{Sak99};
and 741$\pm$37 kpc, based on a study of carbon stars \citep{Dem04}.
\IC\ is highly obscured, with $E_{B-V} = 0.77\pm0.07$ \citep{Rich01}.
With the position of \IC\ on the sky only $\simeq$18$^{\circ}$ apart from
M31, \citet{Bergh00} suggests that it may be a member of
the M31 subgroup. The large number of \ion{H}{ii} regions
\citep{HL90} indicates  that \IC\ is undergoing a massive episode of SF.
Estimates of the intensity of SF were revised upward after many
WR stars were discovered in \IC\ \citep{Mas92, MA95}.
This number of WR stars is remarkable
for a galaxy of its size and is at least a factor of two higher than
the density of such stars seen in any other Local Group galaxy.
This fact led \citet{MA95} to classify \IC\ as a
starburst galaxy, the only
such object in the Local Group. \citet{Rich01} concluded
that \IC\ should  be considered a blue compact dwarf galaxy.

Sixteen PN candidates were identified in \IC\ by \citet{Magr03}
on the basis of both [\ion{O}{iii}] and
H$\alpha$+[\ion{N}{ii}] continuum-subtracted images. No PNe were found very
close to the center of this galaxy, presumably because of the presence of
numerous extended HII regions covering a large fraction of the galaxy's area.
The distribution of m$_{5007}$ magnitudes for all detected PN candidates looks
unusual since the brightest one is $\sim$1\fm7 brighter than the next brightest.
Observations and analysis of three of these PN candidates are presented
below.

The paper is organized as follows. Section~\ref{Obs_red}
gives a description of all observations and the data reduction. Our results
are summarized in Section~\ref{txt:res} and discussed in
Section~\ref{txt:disc}. Throughout the paper, for distance-dependent parameters
we have assumed a distance to \IC\ of 740\,kpc \citep{Dem04}.

\begin{figure}
\centering
\includegraphics[width=8.5cm,angle=0,clip=]{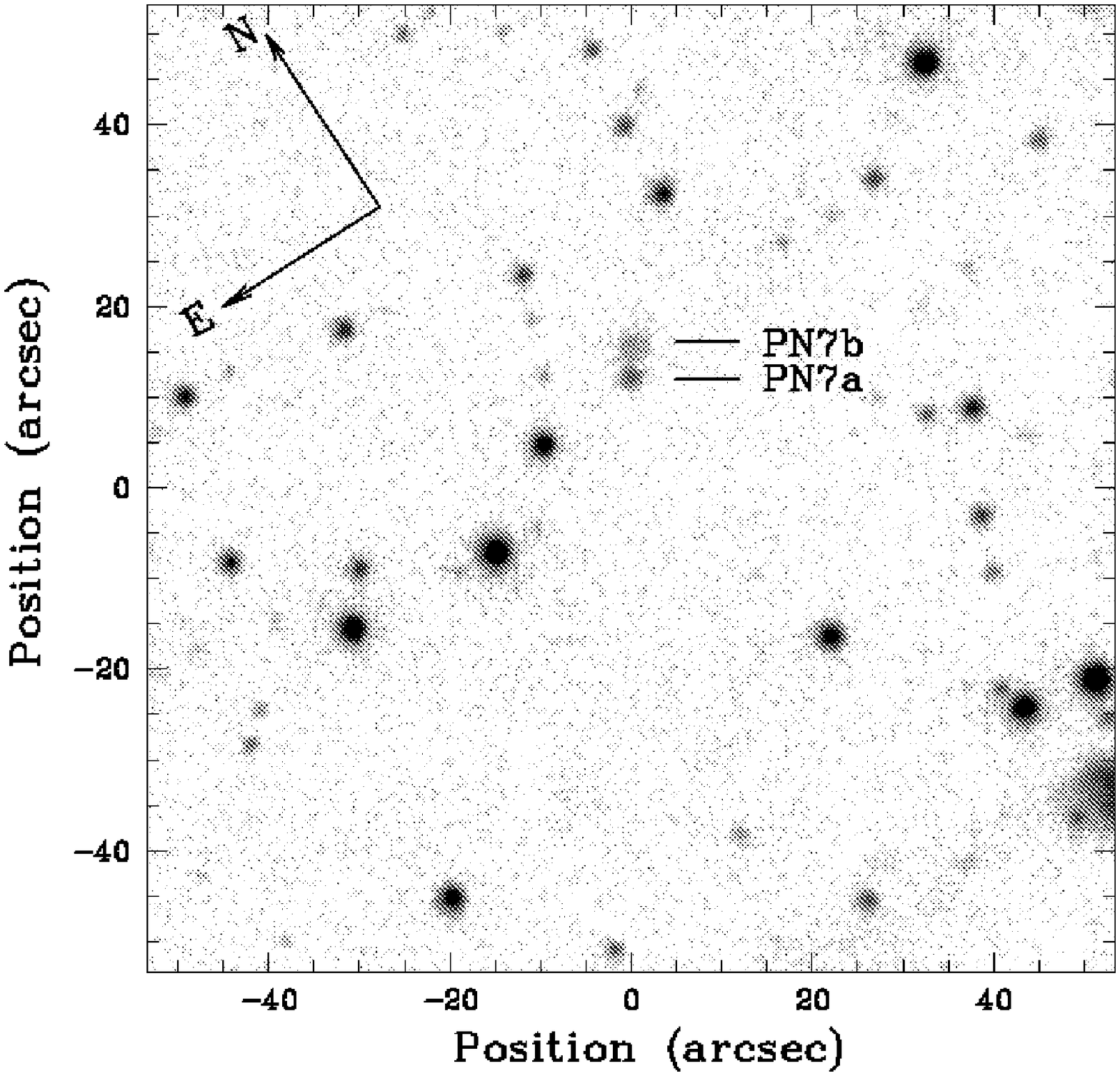}
\includegraphics[width=8.5cm,angle=0,clip=]{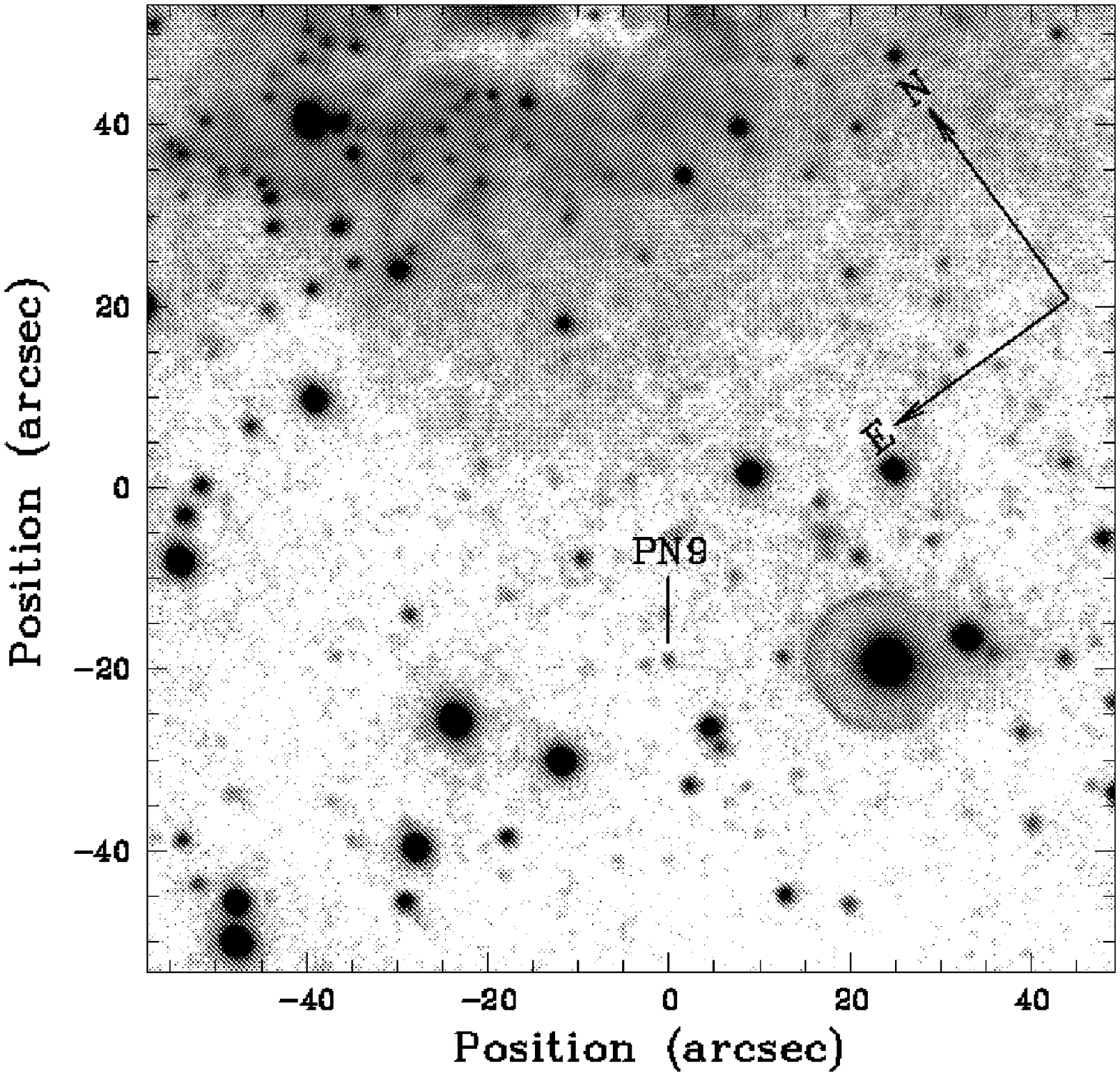}
   \caption{
     Co-added H$\alpha$ images obtained with the 6m telescope for
     PN\,7 (top) and PN\,9 (bottom) candidates in \IC. Image orientations 
     are shown with arrows. The black objects on the
     image correspond to bright sources. 
     The long slits were
     aligned with the Y-axis in this coordinate system, crossing the abcissa at X=0.
     The positions of the observed PN candidates are marked.
     At the assumed distance of 740 Kpc, 1\arcsec\ = 3.6 pc.
	   }
\label{fig:PNe_direct}
\end{figure}

\section{Spectral Observations and Data Reduction}
\label{Obs_red}

Long-slit spectral observations were obtained with the SCORPIO
multi-mode instrument \citep{SCORPIO}
installed in the prime focus of the SAO 6\,m telescope (BTA), during
two nights in November 2004.
The VPH550g grism was used with the 2K$\times$2K CCD detector EEV~42-40,
with an exposed region of 2048$\times$600 px. This gave a wavelength range
$\sim$3500--7500~\AA\ with $\sim$2.0 \AA~pixel$^{-1}$ and FWHM $\sim$12~\AA\
along the dispersion direction.
The scale along the slit was 0\farcs18 pixel$^{-1}$,
with a total length of $\sim$2\arcmin and a slit width of 1\arcsec\ .
The coordinates for the observed PN candidates were taken from \citet{Magr03}.
H$\alpha$ acquisition images were obtained before
the spectroscopic observations in order to select optimal positions for the
slit. In these images, the candidate PN\,7 
(\mbox{$\alpha_{2000.0}$ = 00:20:22.22}, $\delta_{2000.0}$ = 59:20:01.6
from \citet{Magr03} appeared as an elongated object consisting of two
regions: a starlike one (hereafter PN\,7a) and an extended one (hereafter PN\,7b).
The slit was positioned to observe both of these simultaneously.
Candidate PN\,9
(\mbox{$\alpha_{2000.0}$ = 00:20:32.08}, \mbox{$\delta_{2000.0}$ = 59:16:01.6})
is a starlike source.
Finding charts and the slit positions for our observations are shown in
Figure~\ref{fig:PNe_direct}. The exposure times used were 2$\times$15 min
for PN\,7 and 2$\times$20 min for PN\,9, and the candidates were observed at airmasses of
1.8 and 1.1 respectively.
For wavelength calibration, the object spectra were complemented
by reference He--Ne--Ar lamp spectra.
Bias and flat-field images were also acquired for a standard
reduction of 2D spectra. The spectrophotometric standard star Feige~34
\citep{Bohlin96} was observed for flux calibration.

\begin{figure}
\centering
\includegraphics[width=6.3cm,angle=-90,clip=]{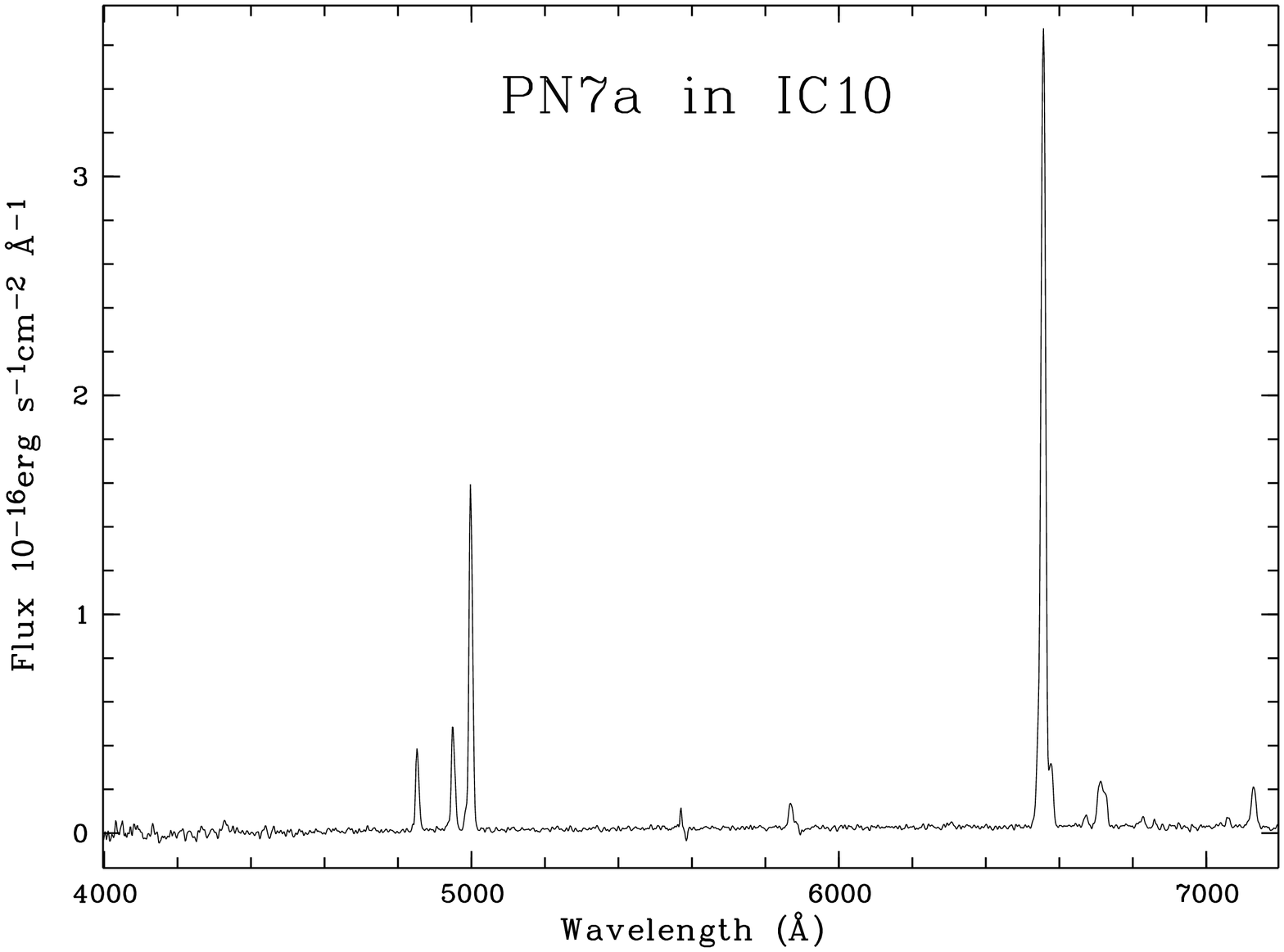}
\includegraphics[width=6.3cm,angle=-90,clip=]{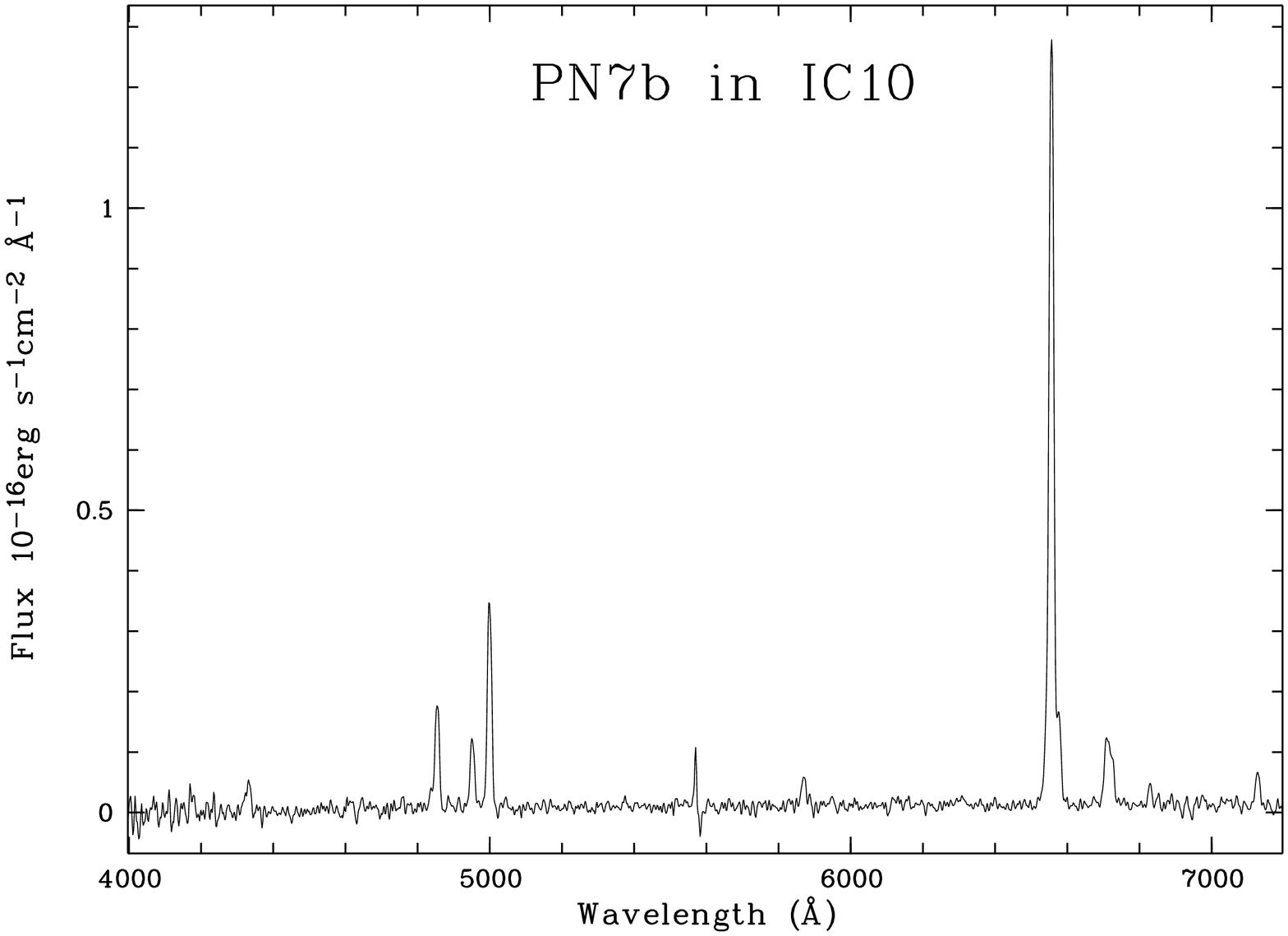}
\includegraphics[width=6.3cm,angle=-90,clip=]{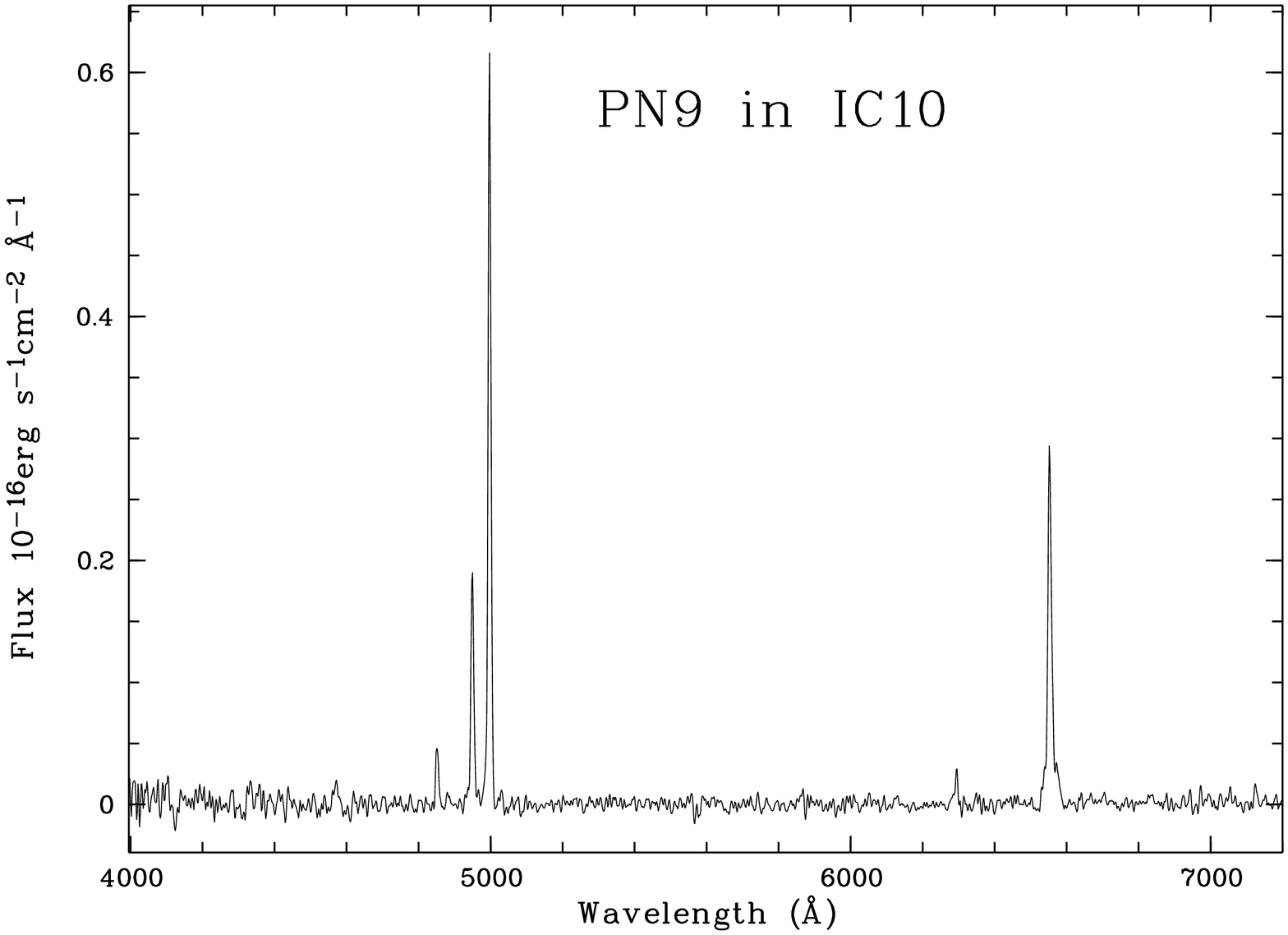}
   \caption{
	    1D spectra of PN candidates PN\,7 and PN\,9 in \IC.
	   }
\label{fig:PNe_spectra}
\end{figure}

Reduction of all data was performed using the standard reduction systems
MIDAS\footnote{MIDAS is the acronym for the European Southern Observatory
package -- Munich Image Data Analysis System.} and IRAF\footnote[2]{The
Image Reduction and Analysis Facility is distributed by the National Optical
Astronomy Observatory, which is operated by the Association of Universities
for Research in Astronomy, Inc., under cooperative agreement with the
National Science Foundation.}.
All cosmic ray hits were removed within MIDAS, while the IRAF package CCDRED
was used for bad pixel removal, trimming, bias-dark subtraction, slit profile
and flat-field corrections. To achieve accurate wavelength calibration,
correction for distortion and tilt for each frame,
sky subtraction and the correction for atmospheric extinction, the IRAF package
LONGSLIT was used. Then, using the data on the spectrophotometric standard
star, each 2D spectrum was transformed to absolute fluxes and one-dimensional
spectra were extracted to obtain the total observed emission line fluxes.

All emission lines were measured using the MIDAS programs described
by \citet{SHOC}. Briefly, they measure the continuum with the
method of \citet{SKL96}, derive robust noise
estimates, and fit resolved lines with a single Gaussian superimposed on the
continuum-subtracted spectrum. Some emission lines which were not sufficiently separated
atr the given spectral resolution were fitted simultaneously as a blend of two
or three Gaussian features. They include the H$\alpha$ $\lambda$6563 and
[\ion{N}{ii}] $\lambda\lambda$6548,6584 lines and the [\ion{S}{ii}]
$\lambda\lambda$6716,6731 lines.
The quoted errors of single line intensities include the following
components:
(1) errors related to the Poisson statistics of line photon flux;
(2) the error resulting from the measurement of the underlying continuum,
which gives the dominant contribution to the errors on faint lines;
(3) an additional error related to the goodness of fit for fluxes of blended lines; and
(4) a term related to the uncertainty in the spectral sensitivity curve
(5\% for the presented observations), which constitutes the  main contribution
to the errors of the relative intensities of strong lines.
All these components are added in quadrature, and the total errors have been
propagated to calculate the errors of all subsequently derived parameters.

\begin{table*}
\centering{
\caption{Line intensities of the observed PN candidates in \IC}
\label{t:Intens1}
\begin{tabular}{lrcccccrr} \hline
\rule{0pt}{10pt}
& \MC{2}{c}{PN\,7a}   && \MC{2}{c}{PN\,7b} && \MC{2}{c}{PN\,9}  \\ \cline{2-3} \cline{5-6} \cline{8-9}
\rule{0pt}{10pt}
$\lambda_{0}$(\AA) Ion                    &
$F$($\lambda$)/$F$(H$\beta$)&$I$($\lambda$)/$I$(H$\beta$) &&
$F$($\lambda$)/$F$(H$\beta$)&$I$($\lambda$)/$I$(H$\beta$) &&
$F$($\lambda$)/$F$(H$\beta$)&$I$($\lambda$)/$I$(H$\beta$) \\ \hline

4340\ H$\gamma$\           & 0.25$\pm$0.05 & 0.461$\pm$0.116    && 0.24$\pm$0.05 & 0.37$\pm$0.08    && 0.32$\pm$0.11  & 0.49$\pm$0.15   \\
4861\ H$\beta$\            & 1.00$\pm$0.08 & 1.000$\pm$0.124    && 1.00$\pm$0.10 & 1.00$\pm$0.12    && 1.00$\pm$0.18  & 1.00$\pm$0.19   \\
4959\ [O\ {\sc iii}]\      & 1.46$\pm$0.12 & 1.263$\pm$0.109    && 0.65$\pm$0.07 & 0.58$\pm$0.07    && 4.35$\pm$0.53  & 3.61$\pm$0.65   \\
5007\ [O\ {\sc iii}]\      & 4.84$\pm$0.37 & 3.980$\pm$0.327    && 2.11$\pm$0.19 & 1.83$\pm$0.18    &&14.81$\pm$1.77  &11.85$\pm$1.93   \\
5876\ He\ {\sc i}\         & 0.36$\pm$0.04 & 0.121$\pm$0.016    && 0.26$\pm$0.05 & 0.13$\pm$0.03    && \MC{1}{c}{---} & \MC{1}{c}{---}  \\
6548\ [N\ {\sc ii}]\       & 0.37$\pm$0.04 & 0.069$\pm$0.009    && 0.31$\pm$0.05 & 0.11$\pm$0.02    && 0.35$\pm$0.12  & 0.11$\pm$0.04   \\
6563\ H$\alpha$\           &14.67$\pm$1.11 & 2.763$\pm$0.404    && 8.46$\pm$0.75 & 2.91$\pm$0.30    && 9.73$\pm$1.18  & 2.91$\pm$0.44   \\
6584\ [N\ {\sc ii}]\       & 1.13$\pm$0.14 & 0.209$\pm$0.031    && 0.97$\pm$0.10 & 0.33$\pm$0.04    && 1.22$\pm$0.23  & 0.36$\pm$0.08   \\
6678\ He\ {\sc i}\         & 0.12$\pm$0.02 & 0.021$\pm$0.004    && ---           & ---              && \MC{1}{c}{---} & \MC{1}{c}{---}  \\
6717\ [S\ {\sc ii}]\       & 0.80$\pm$0.07 & 0.133$\pm$0.016    && 0.71$\pm$0.08 & 0.23$\pm$0.03    && \MC{1}{c}{---} & \MC{1}{c}{---}  \\
6731\ [S\ {\sc ii}]\       & 0.58$\pm$0.06 & 0.096$\pm$0.013    && 0.49$\pm$0.06 & 0.15$\pm$0.02    && \MC{1}{c}{---} & \MC{1}{c}{---}  \\
7136\ [Ar\ {\sc iii}]\     & 0.66$\pm$0.06 & 0.082$\pm$0.010    && 0.32$\pm$0.04 & 0.08$\pm$0.01    && \MC{1}{c}{---} & \MC{1}{c}{---}  \\
\\%
C(H$\beta$)\ dex           & \MC {2}{c}{$\le$2.10$\pm$0.04$^a$} && \MC {2}{c}{$\le$1.35$\pm$0.07$^a$}&&\MC {2}{c}{1.44$\pm$0.12}  \\
F(H$\beta$)$^b$\           & \MC {2}{c}{7.50}          &         & \MC {2}{c}{4.7}           &       & \MC {2}{c}{0.66}           \\
EW(abs)\ \AA\              & \MC {2}{c}{4.00}          &         & \MC {2}{c}{4.30}          &       & \MC {2}{c}{---}            \\
EW(H$\beta$)\ \AA\         & \MC {2}{c}{ 140$\pm$4}    &         & \MC {2}{c}{ 142$\pm$7}    &       & \MC {2}{c}{---}            \\
$N_{\rm e}$(SII)(cm$^{-3}$)& \MC {2}{c}{40$\pm$60}     &         & \MC {2}{c}{ $\le$10}      &       & \MC {2}{c}{---}            \\
Rad. vel.\ \kms\           & \MC {2}{c}{$-$361$\pm$25} &         & \MC {2}{c}{$-$355$\pm$34} &       & \MC {2}{c}{$-$391$\pm$43}  \\
E$_{B-V}$ \ mag            & \MC {2}{c}{$\le$1.43$\pm$0.03} &    & \MC {2}{c}{$\le$0.92$\pm$0.05} &  & \MC {2}{c}{0.98$\pm$0.08}  \\
A$_{5007}$ \ mag           & \MC {2}{c}{$\le$5.01$\pm$0.10} &    & \MC {2}{c}{$\le$3.22$\pm$0.17} &  & \MC {2}{c}{3.43$\pm$0.29}  \\
\hline
\MC{9}{l}{$^a$ Note: PN 7a and 7b were not observed at the parallactic angle. See Section~\ref{txt:res} for more details.}\\
\MC{9}{l}{$^b$ Observed flux in units of 10$^{-16}$ ergs\ s$^{-1}$cm$^{-2}$.}\\
\end{tabular}
 }
\end{table*}

\section{Results}
\label{txt:res}

The relative intensities (normalised to I(H$\beta$)) of all measured emission lines,
as well as the
derived extinction coefficient C(H$\beta$), the equivalent widths (EWs)
of Balmer absorption lines, the measured flux of the H$\beta$ emission line,
the measured heliocentric radial velocity, the electron density N$_{\rm e}$,
the calculated extinction $E_{B-V} = 0.68 \cdot C(H\beta)$,
and the extinction value near $\lambda$5007~\AA, $A_{5007} = 3.5 \cdot E_{B-V}$
\citep{CCM89} are given in Table~\ref{t:Intens1}.
The final reduced 1D spectra are shown in Figure~\ref{fig:PNe_spectra}.
The obtained values of $C$(H$\beta$) in all spectra are high and
correspond respectively to the high derived values of $E_{B-V}$.
Assuming a the Milky Way foreground extinction of $E_{B-V} \simeq 0.77$
in the direction of this galaxy \citep{Rich01},
the internal extinction values $A_{5007}$ vary in the range of 0\fm5 to 2\fm3.
The latter implies that \IC\ contains a considerable amount of internal dust,
a conclusion supported by many previous studies.
Because candidate PN\,7 was observed at high airmass (1.8)
but not at the parallactic angle,
it is possible  that some flux may have been lost 
in the blue and/or red parts of the spectra.
As H$\alpha$ images were used to place objects on the slit, we most likely
lost flux in the blue. Based on \citet{Fil82}, we estimate that we might have lost
$\sim 30\%$ in the spectral region of H$\beta$ and $\sim$50--60\%
in the spectral region of H$\gamma$.
For this reason, the extinction values found for PN\,7 and shown
in Table~\ref{t:Intens1} should be considered upper limits.
Our measured total [\ion{O}{iii}] $\lambda$5007 emission line fluxes
for both observed candidates are about 20-30\%\ lower than those presented by \citet{Magr03}.
This is probably due to a combination of non-photometric conditions
during our observations and the non-parallactic observation angle used for PN\,7.
Fortunately, the uncertain extinction determination for PN\,7
does not affect the results presented in Section~\ref{txt:disc}.

The derived radial velocities of PN\,7a and PN\,7b are close to each other, 
supporting the idea that both objects belong to the same complex. Their
velocities are also close to the optical velocity of \IC\, V$_{hel}$ = $-$348 km s$^{-1}$
\citep{Hu99}.
\citet{SS89} found that \ion{H}{i} velocities in \IC\ range from $-$300 to $-$400
km~s$^{-1}$. Comparison of PN\,9's position and its optical radial
velocity with the 2D \ion{H}{i} velocity distribution
\citep[Figure~7 from][]{SS89} shows agreement to within the uncertainties of
our velocity measurements. Thus, our radial velocities for both PN\,7 and PN\,9
support the idea that the observed PN candidates indeed belong to \IC.

\begin{figure}
\centering
\includegraphics[width=7.5cm,angle=-90,clip=]{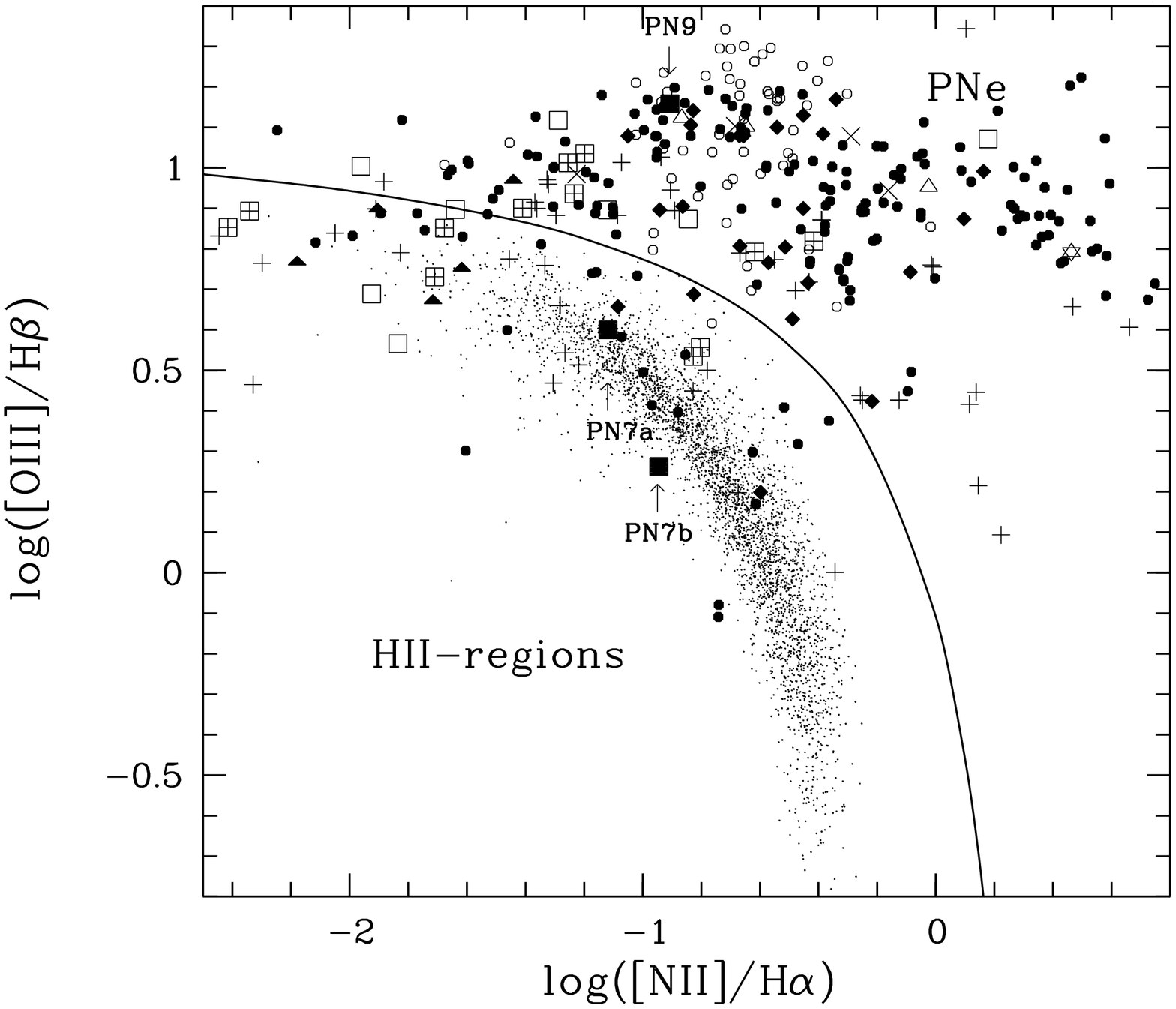}
\includegraphics[width=7.5cm,angle=-90,clip=]{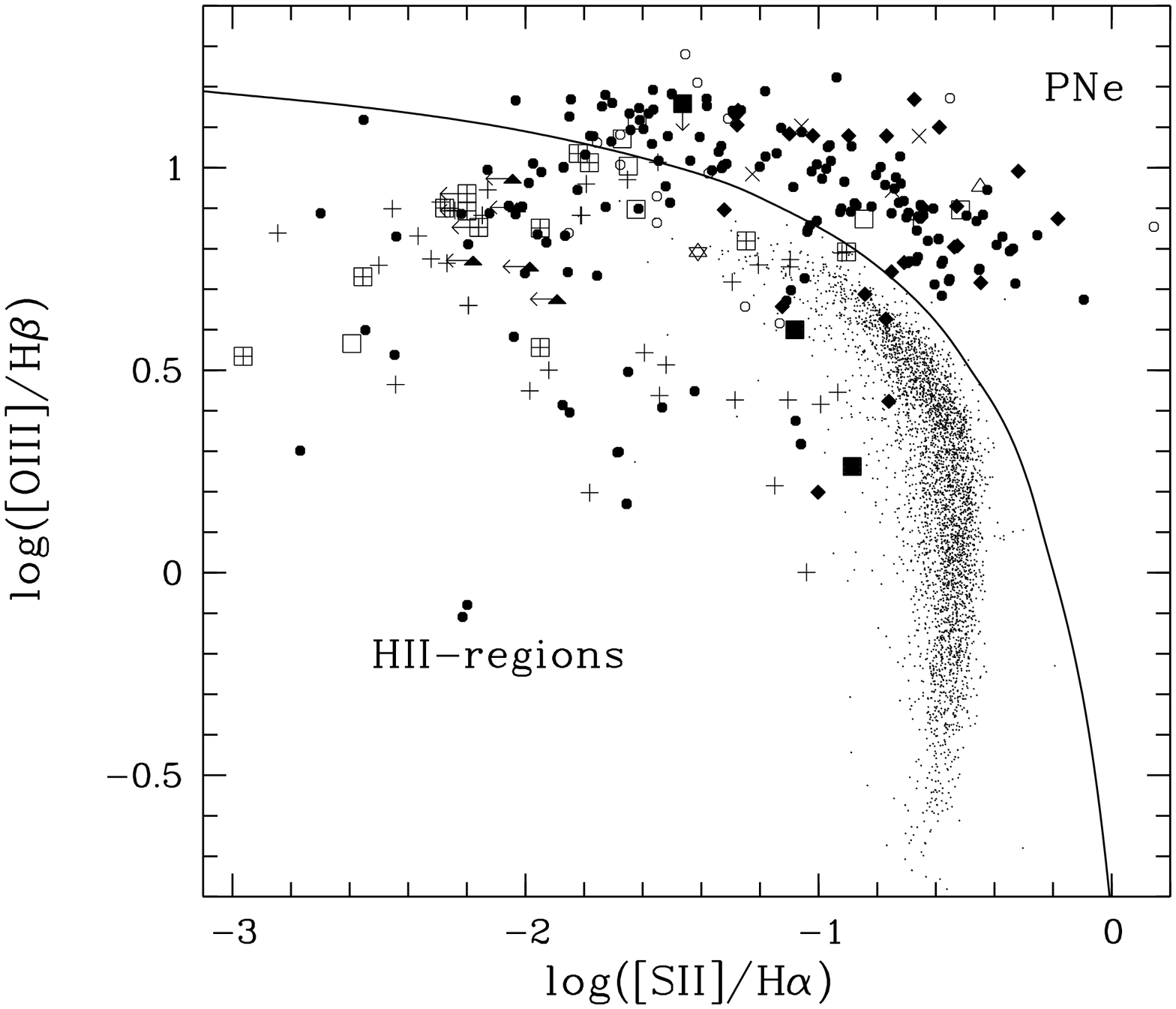}
   \caption{
 Classification diagrams for PNe/\ion{H}{ii} galaxy separation.
 The SDSS DR1 strong emission-line  \ion{H}{ii}-galaxies from
 \citet{SHOC} are plotted as dots.
 The data for extragalactic PNe are overplotted  with different symbols:
 from M~31 as empty circles,
 from NGC~205 and M~32 as empty triangles,
 from Fornax, Sagittarius and NGC~147 as crossed squares,
 from NGC~5128 as crosses (x),
 from the SMC as pluses (+),
 from the LMC as filled diamonds,
 from Leo~A and NGC~6822 as empty squares,
 from Sextans~A as a star,
 from Sextans~B as filled triangles,
 from M~33 as filled lozenges and
 the current data for \IC\ as filled squares.
 The positions of the observed \IC\ PN candidates are labeled
 in the upper panel.
 The points representing upper limits are shown with arrows in the lower panel.
 The solid line shows the models from
 \citet{Kewley01} used for AGN/\ion{H}{ii} galaxy
 separation.
	   }
\label{fig:PNe_class}
\end{figure}

\begin{figure}
\centering
\includegraphics[width=7.5cm,angle=-90,clip=]{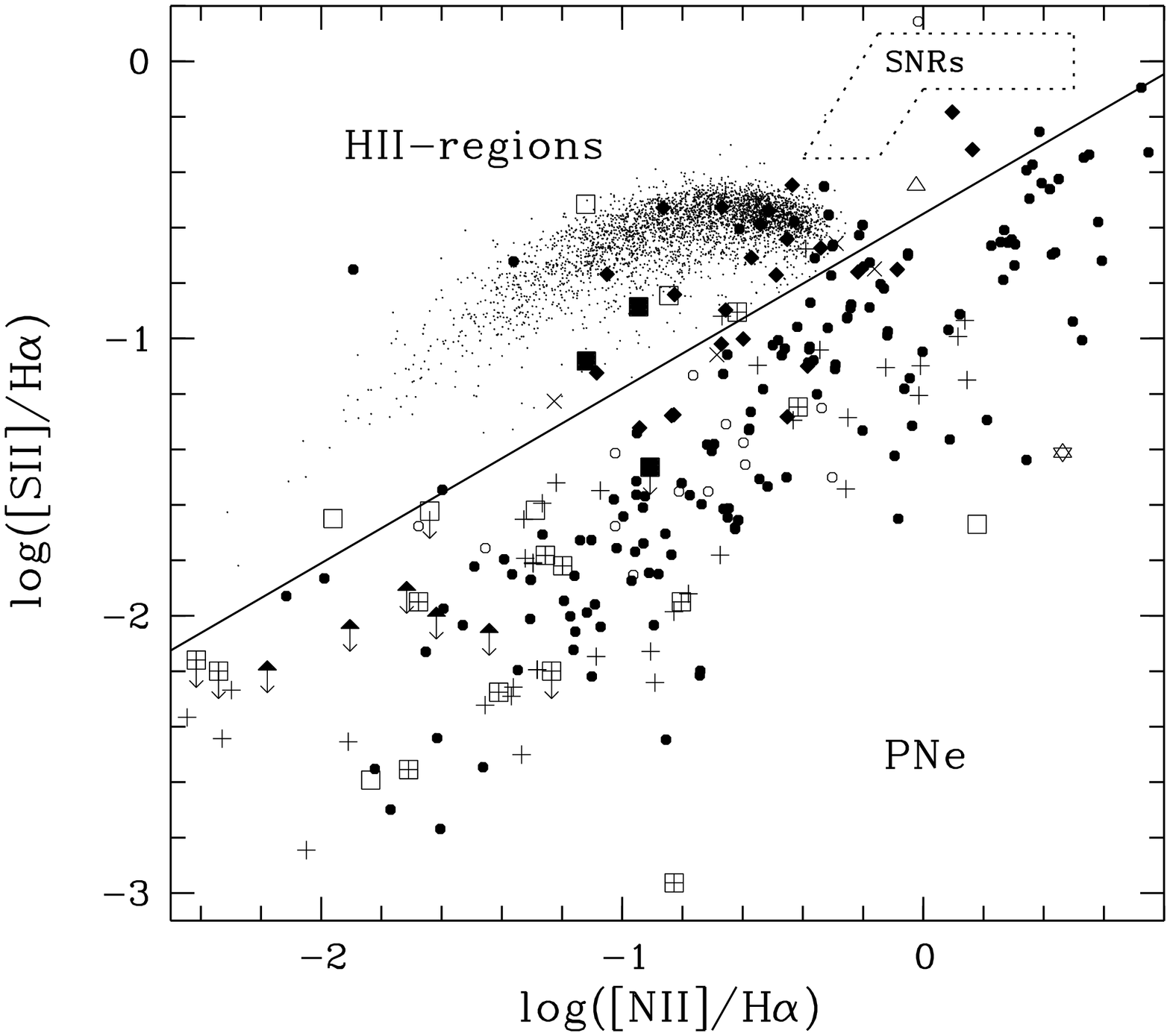}
\includegraphics[width=7.5cm,angle=-90,clip=]{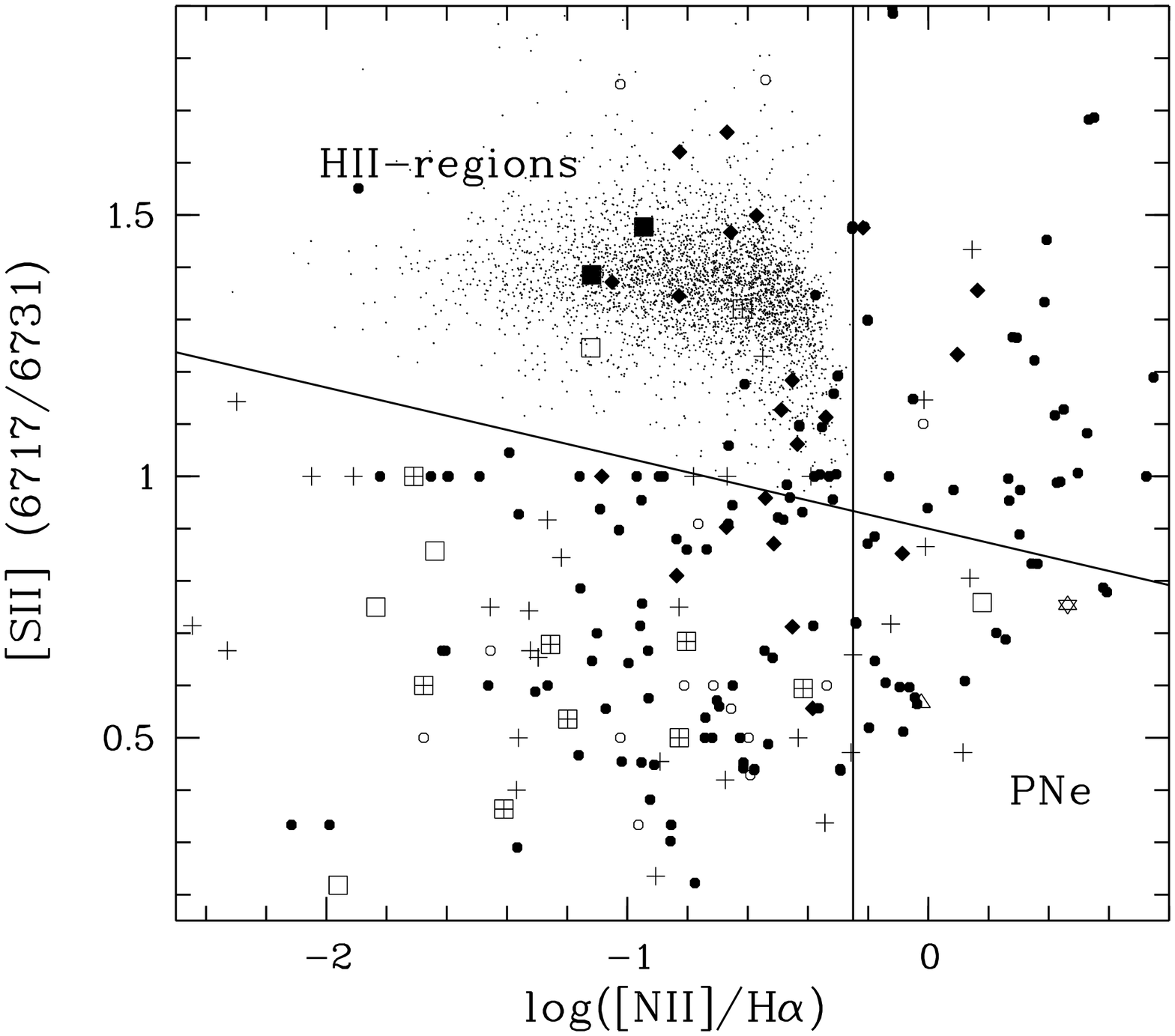}
   \caption{
 Classification diagrams for PNe/\ion{H}{ii} galaxy separation.
 The symbols are the same as in Figure~\ref{fig:PNe_class}.
 The points representing upper limits are shown with downward arrows.
 The solid lines show graphically the  criteria for PNe/\ion{H}{ii}
 separation that were used in this work. See Section~\ref{txt:sel} for more details.
	   }
\label{fig:PNe_class1}
\end{figure}

\section{Discussion}
\label{txt:disc}

\subsection{PN or compact \ion{H}{\sc ii} region?}
\label{txt:sel}

Most extragalactic PN candidates are point-like sources. The task
of distinguishing between PNe and compact \ion{H}{ii} regions in dwarf irregular
and spiral galaxies is extremely difficult and cannot be accomplished with
morphological criteria alone. Usually, the line ratio
\mbox{R=I([\ion{O}{iii}] $\lambda$5007)/I(H$\alpha$ + [\ion{N}{ii}] $\lambda\lambda$6548,6584)}
is used in searches for PN candidates with narrow-band imaging
in H$\alpha$ and [\ion{O}{iii}] $\lambda$5007 lines.
A criterion of $\rm R > 1.6$ for PN candidates was adopted by \citet{Ciar02} after
the analysis of a large sample of extragalactic PNe. However, it is well
known that the ratio $\rm R$ is a function of the absolute magnitude of PNe,
hence many extragalactic PN candidates can have $\rm R < 1.6$
\citet[see, e.g.,  the list of PN candidates from][]{Magr03, Magr05}.
Definitive classification/separation can only be performed with
classification/diagnostic diagrams, after candidate spectra have been
obtained. PNe spectra have the same emission lines as \ion{H}{ii} regions,
but the central stars of PNe are hotter than the OB stars in \ion{H}{ii} regions,
and the line intensity ratios are different for this reason. In addition,
the electron densities in most PNe are higher than those in \ion{H}{ii}
regions. Empirical emission line diagrams are often used both to
distinguish between different classes of ionized nebulae in the Galaxy
\citep{SD76, SMB77, Gar91, RT02, M33, RT06}
and different types of emission-line galaxies
\citep[ELGs, e.g.,][]{H80, Baldwin81, Veilleux87, Ugr99}.
Unfortunately, unlike ELG classification based on such diagrams,
to our knowledge there are no published
quantitative criteria for distinguishing  PNe from  \ion{H}{ii} regions.
Analyses using the $\rm [\ion{O}{iii}] \lambda5007/H\beta$ ratio for
classification purposes are absent from the literature as well, even though
[\ion{O}{iii}]$\lambda$5007 is the strongest line in most PN spectra.

To fill this gap and to derive quantitative criteria for the objective
spectral classification of PN candidates, we first used a subsample of
spectra of \ion{H}{ii} regions (\ion{H}{ii} galaxies) from the SDSS database
\citep{SHOC}. Their line intensity ratios are shown
in Figures~\ref{fig:PNe_class} and \ref{fig:PNe_class1}.
These data cover very large ranges for a variety of emission line intensity ratios
in  \ion{H}{ii} galaxies of different excitation types, and therefore can be
used to define their loci properly. Secondly, we compiled from the
literature a sample of extragalactic PNe with published spectra or emission
line measurements. Data from the following galaxies were included:
M~31 \citep{JF86, JC99, Ket05}, M~33 \citep{M33},
LMC and SMC \citep{All83, MBC88, BL89, MD91a, MD91b, Vas92, JK93, Do93, LD06},
Sextans~A and Sextans~B \citep{Sextans, SextB},
Leo~A \citep{Ket05}, NGC~6822 \citep{RM07},
Fornax \citep{KGPP04, Fornax},
Sagittarius \citep{Walsh, Zijetal06}, NGC~147 \citep{Gon07},
M~32 \citep{JFJ79},
NGC~205 \citep{RM95}, and NGC~5128 \citep{Walsh99}.
In cases for which measurements of [\ion{S}{ii}] lines in PNe and PN candidates were not available, and for
which we have data (PNe
in Sextans~B and PN\,7 in \IC), $2\sigma$ upper
limits were calculated. These upper limits are shown as well with arrows.
Our data for \IC\ were not used for criteria selection.

In total, we collected a sample of 259 different extragalactic PNe with
measurements of all the emission line intensities we are considering:
[\ion{O}{iii}] $\lambda$5007, {\rm H$\beta$}, {\rm H$\alpha$},
[\ion{N}{ii}] $\lambda$6584, and [\ion{S}{ii}] $\lambda\lambda$6716,6731.
We did not require separate measurements of the [\ion{S}{ii}] $\lambda\lambda$6716,6731
lines for this sample, which we used for calculation of the overall selection efficiency.
A subsample of 227 extragalactic PNe with resolved measurements of
[\ion{S}{ii}] $\lambda\lambda$6716,6731  was used
to estimate the selection efficiency for criteria involving the use of the
sulphur lines.

Our analysis has shown that it is possible to classify/separate almost all
collected extragalactic PNe,
if at least two (main) diagrams are used:
\mbox{log([\ion{O}{iii}] $\lambda$5007/H$\beta$)}  vs.\,
\mbox{log([\ion{N}{ii}] $\lambda$6584/H$\alpha$)} and
\mbox{log([\ion{S}{ii}] $\lambda\lambda$6716,6731/H$\alpha$)}  vs.\,
\mbox{log([\ion{N}{ii}] $\lambda$6584/H$\alpha$)}.
However, it is worth noting that most of these PNe can easily be separated
with only the first diagram, which utilizes the strongest emission lines
easily detectable in both PNe and \ion{H}{ii} regions.
It is clearly seen in the top panel of Figure~\ref{fig:PNe_class} that
most PNe are located in the same area as AGNs and can be easily distinguished
with the model track of \citet{Kewley01}, often used
for AGN/\ion{H}{ii} segregation \citep[see, e.g.,][]{SHOC}.
The demarcation line from this model is shown as a solid line in
Figure~\ref{fig:PNe_class}. We use this to construct the first quantitative
criterion for the PNe selection:
\begin{equation}
\label{equ:1}
{\rm \log{([\ion{O}{iii}] /H\beta)} \ge (0.61/\log{([\ion{N}{ii}] /H\alpha)} - 0.47) + 1.19}
\end{equation}
This allows us to identify 76\%\ of all objects from the compiled
list as PNe. However, the subsample of PN in Leo~A, four PNe in Sextans~B, the
PN\,7a and PN\,7b candidates in \IC\ and some PNe candidates from M33, LMC and
SMC form a locus in the area of \ion{H}{ii} regions, so that other
classification diagrams must be used to separate them.
The use of the diagram
\mbox{log([\ion{O}{iii}] $\lambda$5007/H$\beta$)}  vs.\,
\mbox{log([\ion{S}{ii}] $\lambda\lambda$6716,6731/H$\alpha$)}
(bottom panel of Figure~\ref{fig:PNe_class}) and the \citet{Kewley01}
model's demarcation line
\begin{equation}
\label{equ:1a}
{\rm \log{([\ion{O}{iii}] /H\beta)} \ge (0.72/\log{([\ion{S}{ii}] /H\alpha)} - 0.32) + 1.30}
\end{equation}
does not help, since more PNe appear in the region of \ion{H}{ii}
compared with those found in the diagram
\mbox{log([\ion{O}{iii}] $\lambda$5007/H$\beta$)}  vs.\,
\mbox{log([\ion{N}{ii}] $\lambda$6584/H$\alpha$)}.
The classification efficiency of criterion (\ref{equ:1a}) for our compiled PNe
sample is only 62\%. Thus we did not use the classification diagram
\mbox{log([\ion{O}{iii}] $\lambda$5007/H$\beta$)}  vs.\,
\mbox{log([\ion{S}{ii}] $\lambda\lambda$6716,6731/H$\alpha$)}
in any further analysis.

The separation of supernova remnants (SNRs) is outside the scope
of the current study. However,
we suggest that most SNRs can also be distinguished
with the use of the diagram
\mbox{log([\ion{O}{iii}] $\lambda$5007/H$\beta$)}
vs.\, \mbox{log([\ion{N}{ii}] $\lambda$6584/H$\alpha$)},
the model lines from \citet{Kewley01}, and the addition of
model lines for LINER selection from \citet{Veilleux87}
and \citet{Baldwin81}, as was demonstrated in \citet{SHOC}.
The latter is possible since
both LINERs and SNRs have significant emission contributions from shock
excitation.
All such objects are located below the line corresponding to
criterion (\ref{equ:1}). The locus of SNRs found in M~33
\citep{M33} supports this conclusion.

The diagram \mbox{log([\ion{S}{ii}] $\lambda\lambda$6716,6731/H$\alpha$)}
vs.\, \mbox{log([\ion{N}{ii}] $\lambda$6584/H$\alpha$)}
(shown in the top panel of Figure~\ref{fig:PNe_class1}), can be used
as an additional diagnostic for the identification of PNe.
From the analysis of our compiled PN data
we have found that, with the use of only the criterion:
\begin{equation}
\label{equ:2}
{\rm \log{([\ion{S}{ii}] /H\alpha)} \le 0.63 \cdot \log{([\ion{N}{ii}] /H\alpha)} - 0.55}
\end{equation}
83\%\ of our PNe can be recovered.
However, with the combined use of criteria (\ref{equ:2}) and
(\ref{equ:1}), 99\%\ of PNe  can be distinguished
from \ion{H}{ii} regions (i.e., all but two cases).
Any analysed object is classified as a PN if it is located in the PNe
locus according to at least of one of the cited criteria.
It should be noted that, for many extragalactic PNe the detection of
[\ion{S}{ii}] $\lambda\lambda$6716,6731 lines is difficult, since the ratio
I([\ion{S}{ii}] $\lambda\lambda$6716,6731)/I(H$\beta$) is typically
$\sim$0.5--5\%. However, upper limits for this line ratio can be
successfully used in this situation, similar to the cases of the PNe from Sextans~B
and the candidate PN\,9 in \IC.
This second criterion is also very useful for rejecting possible SNRs,
which fill the locus shown with the dashed line
in the top panel of Figure~\ref{fig:PNe_class1}.

The last classification diagram we considered is the relation between
the flux ratio of the two lines of the [\ion{S}{ii}] $\lambda$6716/$\lambda$6731
doublet  and that of the flux ratio of the [\ion{N}{ii}] $\lambda$6584 and H$\alpha$
lines:
[\ion{S}{ii}] $\lambda$6716/$\lambda$6731 vs.\,
\mbox{log([\ion{N}{ii}] $\lambda$6584/H$\alpha$)}.
This is shown in the bottom panel of Figure~\ref{fig:PNe_class1}.
The criteria below
\begin{equation}
\label{equ:3}
{\rm [\ion{S}{ii}] \lambda6716/\lambda6731 \le -0.135 \cdot \log{([\ion{N}{ii}] /H\alpha)} + 0.90}  \\
\end{equation}
\begin{equation}
\label{equ:3a}
{\rm \log{([\ion{N}{ii}] /H\alpha)} \ge -0.25}
\end{equation}
select 81\%\ of PNe that have their
[\ion{S}{ii}] $\lambda\lambda$6716,6731 lines measured separately.
With a combination of criteria (\ref{equ:1}) and (\ref{equ:2}), only one more
PN is selected and the total selection efficiency reaches 99.6\%.
Just one PN from our compiled list, MCMP49 in M~33 \citep{M33},
cannot be identified as a genuine PN using all the diagrams.

\begin{figure}
\centering
\includegraphics[width=7.5cm,angle=-90,clip=]{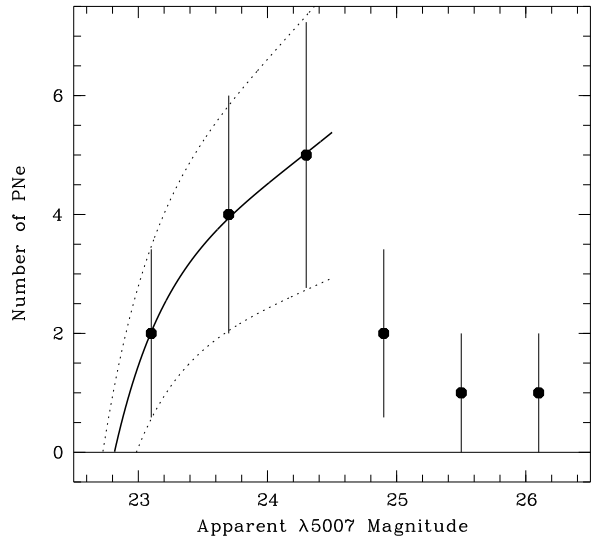}
   \caption{
PNLF for the PN candidates in \IC, excluding PN\,7.
The data have been binned in 0\fm6 intervals.
The solid curve represent the best-fitting ``universal'' PNLF \citep{CJFN89}
obtained for the data points.
The dotted curves represent the best-fitting ``universal'' PNLFs
obtained for the data points with $\pm1\sigma$ errors.
	   }
\label{fig:PNLF_IC10}
\end{figure}

\subsection{On the brightest PN in \IC\ and its distance}
\label{txt:bright}

The \IC\ candidate PN\,9 observed in this work can be confidently
classified as a true PN according to both criteria (\ref{equ:1}) and
(\ref{equ:2}). However, according to
all classification diagrams shown in Figures~\ref{fig:PNe_class} and
\ref{fig:PNe_class1}, \IC\ candidates PN\,7a and PN\,7b,  cannot be selected as
PNe and for this reason were finally classified as compact \ion{H}{ii} regions.
An additional reason to believe that PN\,7a and PN\,7b are \ion{H}{ii} regions
is the detection of continuum in their spectra (see
Figure~\ref{fig:PNe_spectra}).

The absolute magnitude cutoff of the PNe [\ion{O}{iii}] $\lambda$5007
luminosity function (PNLF) is considered to be constant in large galaxies
with a large population of PNe, and is equal to
$M_{*5007} = $-$4\fm47$ \citep{Ciar02}.
For dwarf galaxies, this cutoff is likely shifted towards
the fainter magnitudes as a function of the galaxy mass and the population
size \citep{Men93}.
\citet{Dop92} also examined
the effects of metallicity on the luminosity of the brightest PN. According
to their models, the variation of the cutoff magnitude $M_{*5007}$  with
oxygen abundance is described by the relation
\begin{equation}
\label{equ:abun}
{\rm \Delta M_{*5007} = 0.928 \cdot [O/H]^2 + 0.225 \cdot [O/H] + 0.014},
\end{equation}
where [O/H] is the system's logarithmic oxygen abundance referenced
to the solar value 12+log(O/H) = 8.87 \citep{Anders89}.
\citet{Ciar02} studied the accuracy of $M_{*5007}$ and
found that $M_{*5007} = $-$4\fm47\pm0.05$ after the correction for metallicity
using equation (\ref{equ:abun}).

The brightest PNe absolute magnitudes should be reasonably close
to M$_{*5007}$.
With PN\,7, the brightest candidate in \IC\ from \citet{Magr03},
M$_{*5007}$ for this galaxy can be calculated as $-5\fm34$,
where for candidate PN\,7 $m_{5007} = 21\fm71$ \citep{Magr03},
$A_{5007} = 2\fm70$, assuming background extinction of $E_{B-V} = 0.77$
\citep{Rich01, H01} and a distance modulus  $\rm (m-M)$ = 24.35 \citep{Dem04}.
Adopting the value of 12+$\log$(O/H) = 8.19,
[O/H] = $-$0.68 and $\Delta M_{IC10,*5007}$ = 0.29,
respectively for \IC\ we expect
the `corrected' cutoff magnitude for the \IC\ PNLF to be $M_{*5007} = $-$4\fm18$.
Thus, the calculated M$_{*5007}$ for PN\,7 appears $\sim$1\fm15 brighter
than the standard value.

After the rejection of the PN\,7 candidate as an \ion{H}{ii} region based on the results of our
observations,
the PNLF for PN candidates in \IC\ is shown in Figure~\ref{fig:PNLF_IC10},
where data from \citet{Magr03}
have been binned into 0\fm6 intervals. Poisson $\pm1\sigma$ error bars
are shown for the detected number of PNe in these bins.
We have fitted this PNLF with a ``universal'' PNLF \citep{CJFN89},
taking only the brightest PNe (first three bins, eleven PNe in total)
into account
and using a uniform foreground extinction of
$E_{B-V} = 0.77$ \citep{Rich01, H01}.
The result of this fitting and Poisson errors are also shown in Figure~\ref{fig:PNLF_IC10}.
With this small sample we obtained the distance modulus
\mbox{$\rm (m-M) = 24.30^{+0.18}_{-0.10}$}  ( including the error in $M_{*5007}$),
corresponding to a distance D = 725$^{+63}_{-33}$ kpc; this is very close to the value
from \citet{Dem04}. This distance modulus is also 
consistent with the recent estimate $\rm (m-M) = 24.48\pm0.18$
in \citet{Vac07}, obtained from the study of individual RGB stars.
With the PN data in hand, a distance to \IC\ of 950 Mpc \citep{MA95, H01} seems implausible.
The PNLF dip after the first three bins is likely a result of
incompleteness, or could arise in a young population
in which central star evolution proceeds very quickly \citep{JDM02}.
Such a dip is also detected in the PNLFs in NGC\,6822 \citep{Leisy05}, the
Small Magellanic Cloud and M\,33 \citep{JDM02,C04}.

Of course, PN candidates, not confirmed PNe, were used for the PNLF calculation
(except in the case of PN\,9, which was confirmed in this paper), making this result
for the \IC\ PNLF distance somewhat preliminary. However,
all the bright PN candidates that were used for the PNLF fitting have
$\rm R > 1.6$ \citep{Magr03} and thus have a very high
probability of being  real PNe.

The calculated distance to \IC\ obviously depends on the
extinction values used; higher extinction values would place \IC\ even closer to us.
The total extinction value  toward \IC\ is a combination of foreground (i.e., Galactic) extinction,
and the extinction internal to \IC\ ; this latter quantity is quite uncertain, and for this reason
only foreground extinction is usually considered (as we have done in this work).
The great advantage of PNLF distance calculations (compared to other methods) is that,
in principle, the total extinction along the line of sight can be calculated
after proper spectral observations of PN candidates
and calculations of C(H$\beta$) for each PN \citep{JWC90}.
For this reason further spectroscopic observations of other PN candidates in \IC\ are
very important.
On the other hand, total extinctions calculated using C(H$\beta$) for
PNe have to be used carefully, since could be strongly affected by extinction in
localised circumstellar material.
For this reason we do not make further use of the extinction we derive from the spectrum of
PN\,9 in \IC; instead, we plan to map the line-of-sight extinction in its neighbourhood through observations of nearby
\ion{H}{ii} regions.

\section{Conclusions}
\label{txt:concl}

In this paper we present the first results of follow-up spectroscopy
of PN candidates in the Local Group starburst galaxy \IC. Based on our data
and the discussion above, we draw the following conclusions:

1. From the obtained spectral data and the emission line diagnostic diagrams
we have found that the brightest candidate PN\,7 from the list
of \citet{Magr03} is not a genuine PN, but rather a close pair of
compact \ion{H}{ii} regions.
The calculated absolute magnitude M$_{*5007}$ for PN\,7 is about 1\fm15 brighter
than the standard maximum value from the PN luminosity function
expected at the approximate distance of \IC.

2. We have found that PN\,9, obtained from the same list of candidate PNe,
is a true PN, and thus is the first confirmed PN in \IC.

3. With the available PN candidate data,
we estimate the PNLF distance to \IC\ as 725$^{+63}_{-33}$ kpc,
a distance modulus of $\rm (m-M) = 24.30^{+0.18}_{-0.10}$.

\section*{Acknowledgments}
The authors thank A.G. Pramskij  and A.V. Moiseev for their help in
observations with SCORPIO.
S.A.P. acknowledges partial support from the Russian state program
"Astronomy".
We thank the anonymous referee for comments and suggestions which
helped to improve the presentation of the manuscript.
This research has made use of the
NASA/IPAC Extragalactic Database (NED) which is operated by the Jet
Propulsion Laboratory, California Institute of Technology, under contract
with the National Aeronautics and Space Administration. We have
also used the Digitized Sky Survey, produced at the Space Telescope Science
Institute under government grant NAG W-2166.

\bsp

\label{lastpage}

\end{document}